\documentclass[prl,aps,floatfix,preprint,showpacs]{revtex4}
\usepackage{epsfig}

\newcommand{\be}{\begin{equation}}
\newcommand{\ee}{\end{equation}}
\newcommand{\bea}{\begin{eqnarray}}
\newcommand{\eea}{\end{eqnarray}}

\def\figone#1#2#3{\begin{figure}
\centering \leavevmode
\epsfxsize=0.8\columnwidth \epsfbox{#1}
\caption{#2 \label{#3}}
\end{figure} }

\begin{document}

\title{Critical Dynamics of Dimers: Implications for the Glass Transition}


\author{Dibyendu Das}

\affiliation{Indian Institue of Technology Bombay,  Powai, India 400076}

\author{Greg Farrell,  Jane' Kondev and Bulbul Chakraborty }

\affiliation{Martin Fisher School of Physics, Brandeis University,
Mailstop 057, Waltham, Massachusetts 02454-9110, USA}

\begin{abstract}

The Adam-Gibbs view of the glass transition relates the relaxation
time to the configurational entropy, which goes continuously to zero at
the so-called Kauzmann temperature. We examine this scenario in the
context of a dimer model with an entropy vanishing phase transition,
and stochastic loop dynamics. We propose a coarse-grained
master equation for the order parameter dynamics which is used to
compute the time-dependent autocorrelation function and the
associated relaxation time.
Using  a combination of exact results, scaling arguments and numerical
diagonalizations of the master equation, we find
non-exponential relaxation and a Vogel-Fulcher divergence of the
relaxation time in the vicinity of the phase transition. Since in the
dimer model the entropy stays finite
all the way to the phase transition point, and then jumps discontinuously
to zero, we demonstrate a clear departure from the Adam-Gibbs scenario.
Dimer coverings are the ``inherent structures'' of the canonical frustrated system,
the triangular Ising antiferromagnet. Therefore, our results provide a
new scenario for the glass transition in supercooled liquids  in terms of
inherent structure dynamics.
\end{abstract}


\maketitle

\section{Introduction}
A variety of systems such as supercooled liquids, colloids,
granular matter and foams,  exhibit a transition from a flowing
fluid phase to a frozen solid phase. Jamming due to spatial
constraints imposed on the elementary constituents of these
materials has been proposed as a possible common cause of this
dynamical arrest \cite{edwards,nagel_jamming,weitz}.

Model systems, such as hard spheres, have an important role to
play in the investigation of such a scenario since they allow for
a precise definition of jamming \cite{torquato}. They are also
useful in elucidating the precise relationship between
thermodynamics and dynamics in materials exhibiting a jammed phase
\cite{krauth}.  The entropy-based Adam-Gibbs
theory\cite{Adam_Gibbs}relates the viscosity (a dynamical
quantity) to the configurational entropy ($S_{conf}$)(a
thermodynamic quantity)  through $\eta = {\eta}_0
\exp(A/TS_{conf})$. The ideal glass transition is associated with
the Kauzmann temperature at which the configurational entropy
vanishes\cite{Gibbs_dimarzio}.  In this paper, we explore the
connection between dynamics and thermodynamics in a lattice model
of dimers with an entropy-vanishing phase transition.

The dimer model is one of the working horses of statistical mechanics.
It provides an example of a jammed system which has the added
advantage of being exactly solvable \cite{nagle_review}. States of
the dimer model are  specified by placing dimers on the bonds of
the lattice so that every lattice site is covered by exactly one dimer;
see Fig~\ref{dimer_fig}. These dimer coverings are ``locally
jammed'' \cite{torquato} as every dimer cannot move to an empty,
neighboring bond, without violating the packing constraint. Moves
that involve {\em loops} of dimers and adjacent empty bonds, on
the other hand, are allowed. An example of such a move for the
hexagonal lattice involving an elementary plaquette is shown in
Fig.~\ref{dimer_fig}a. Stochastic dynamics of the dimer model on
the square lattice based on these elementary moves were considered
by Henley \cite{clh}.

Most states of the dimer model allow for
elementary moves; an example  of one which does not is shown in
Fig.~\ref{dimer_fig}b. The smallest move in this case involves a
system spanning loop, and we call this state ``maximally jammed''.
If we define an energy functional on the space of dimer coverings
which favors the maximally jammed state, a transition into this
state can be affected as the temperature is lowered. The central question
we address in this paper is: {\em What happens to relaxation time
scales of the dimer model as the transition to the maximally
jammed state is approached?} We will show that the relaxation is
dominated by entropy barriers and is sensitive to
equilibrium fluctuations near the phase transition point.

\figone{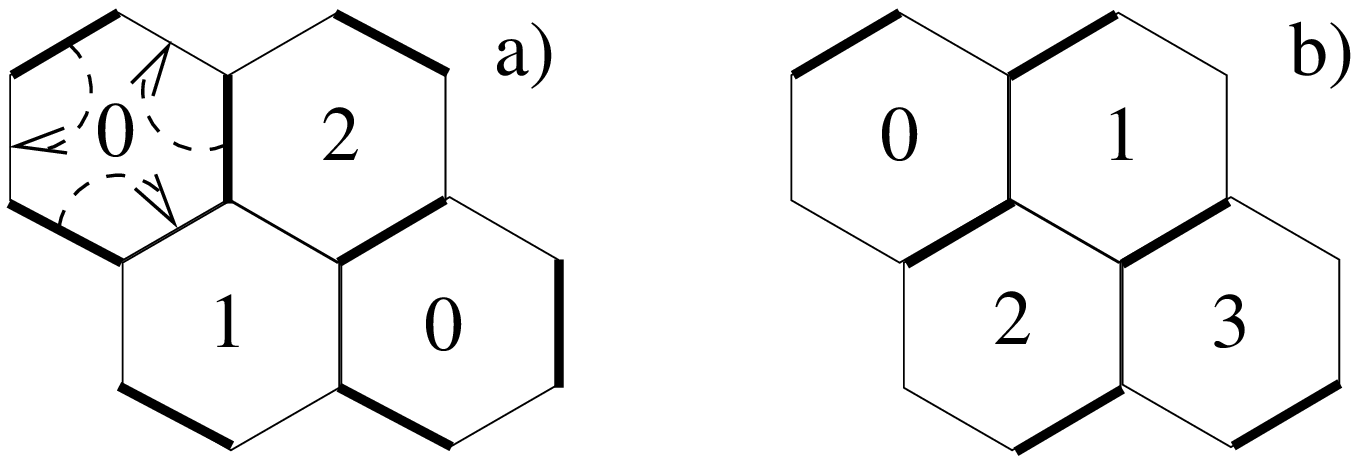}{(a) Dimer covering of the honeycomb
lattice with an elementary loop update indicated by the arrows.
The numbers are the heights of the equivalent interface.  (b) An
ordered, maximally jammed dimer covering; the equivalent interface
is tilted with maximum slope.}{dimer_fig}

We consider an energy functional that exhibits a continuous transition to
the maximally jammed state along a metastable line. We find a strong
departure from the canonical critical-slowing-down scenario
\cite{halperin}, which we attribute to the presence of entropy barriers. Barriers
can be traced directly to the non-local nature of the dynamical
moves allowed by the jammed states. The longest relaxation time-scale is
found to diverge exponentially following a Vogel-Fulcher-like
form. This is reminiscent of what is observed in fragile glass
formers \cite{angel_review}. 
Exponential time-scale divergence  (activated scaling) is associated with critical points in models with quenched 
disorder\cite{fisherscaling}  and it has been argued that real glasses belong to the universality class of random
Hamiltonians with such exponential divergences\cite{parisi}.   The current model provides an explicit example of a
model without quenched disorder which exhibits activated scaling.  It should be mentioned that the Vogel-Fulcher law has 
been observed in models with entropic barriers \cite{backgammon}, with traps \cite{bouchaud}, 
and within effective medium theory \cite{kumar}, none of which have 
an explicit critical point.

\section{Dimer model}
We consider the  dimer model on the  2-d hexagonal
lattice of linear size $L$, having $2 L^2$ sites and $3 L^2$ bonds,
with periodic boundary conditions \cite{kasteleyn}.
A useful representation of the dimer model is given by the height map which
associates a discrete interface $h(x,y)$ with every dimer covering \cite{blote}.
The heights of the interface are defined on the vertices of the dual
triangular lattice. The height difference $\Delta$ between two nearest neighboring
sites is -2 or +1 depending on whether the bond of the honeycomb lattice
that separates them is occupied by a dimer or not; see Fig.~\ref{dimer_fig}a.
Directions in which the height change is $+\Delta$ are specified by orienting
all the up pointing triangles of the dual lattice clockwise.

The dimer model has an extensive entropy.  The ensemble of equal
weighted dimer coverings maps to a rough surface with a
gradient-square free energy \cite{blote}. Fluctuations of the surface
are entropic in origin. A phase transition can be induced in the dimer
model by including an energy functional which is minimized by a dimer
covering corresponding to a smooth, {\em maximally tilted} surface
which corresponds to the maximally jammed state shown in
Fig.~\ref{dimer_fig}b.

For periodic boundary conditions the tilt vector, $({\Delta}_x h ,
{\Delta}_y h)$, where ${\Delta}_{x,y} h$ is the average height difference
in the $x$ or $y$ direction, has only one independent component
$\rho$ \cite{foot2}.  In terms of $\rho$, the energy functional we consider can be
written as:
\be \beta E({\rho}) = -{{\mu L^2} \over 3} ({1}+{8}{{\rho}}^2) ,
\label{energyF}
\ee
where $\mu$ is a dimensionless coupling, proportional to  inverse temperature ($\beta = 1/kT$),  that drives the transition.

The entropy of the dimer model as a function of $\rho$
was calculated exactly \cite{kasteleyn,adhar}:
\be
S({\rho}) = L^2 \left \{ {{2 \ln2} \over 3} (1-{\rho}) +
{2 \over \pi}\int_{0}^{{\pi \over 3}(1-{\rho})}dx \ln[\cos x] \right \}
\label{entropyD1}
\ee
This function has a maximum at $\rho=0$ which is the equilibrium value at
$\mu=0$. For finite $\mu$ this dimer model was previously considered in
Ref.~\cite{hui}. A dimer model with a similar phase transition but
with an energy functional linear in $\rho$ was solved exactly by
Kasteleyn \cite{kasteleyn}.

In the dimer model with the free energy  ${\beta}F={\beta}E-S$, and  the energy and
entropy given by Eqs.~\ref{energyF} and~\ref{entropyD1}, there is an
interesting phase transition along the metastable line,
when the order parameter is confined to the free energy well around
the zero-tilt state.  Namely, at $\mu_{*} = \pi/(8\sqrt{3})$, the
end-point of the metastable line, the order parameter ${\rho}$ has a
discontinuous jump from $0$ to $1$, characteristic of a first-order
transition.  At the same time, as $\mu_{*}$ is approached from below,
fluctuations of ${\rho}$ around $0$ diverge, as would be expected at a
critical point.  This transition was discussed in detail in Ref.~\cite{hui}.
In this paper we investigate the dynamics of the dimer model near this
phase transition point.

\section{Coarse-grained dynamics}
As mentioned in the introduction, the hard constraint of
no overlapping of dimers, gives rise to nonlocal dynamics. We consider stochastic,
Monte-Carlo dynamics based on loop updates with
loops of arbitrary size; a concrete implementation
is given in Ref.~\cite{krauth1}. Since  we
take periodic boundary conditions, loops with different winding
numbers can be formed. We restrict loop updates to loops with winding
numbers $(0,0), (1,0)$ and $(0,1)$, only.  The microscopic transition rates for
loop updates are given by Metropolis rules that follow from the energy function,
Eq.~\ref{energyF}.

Given the microscopic loop dynamics, which satisfy conditions of ergodicity and
detailed balance, we ask what are the coarse-grained dynamics of the order
parameter, $\rho$.  Since the energy function in Eq.~\ref{energyF} depend on the global tilt
$\rho$ only, it follows that all updates of topologically trivial
loops (i.e.~those with $(0,0)$ winding number) have $\Delta E =
0$. Only when system spanning loops with nonzero winding numbers are
updated does the energy of the state change.
This feature naturally leads to fast and slow processes in the
Monte-Carlo dynamics. On a faster time scale, non-winding loops are
updated with no effect on the overall tilt of the surface, while on a
much slower time scale, winding loops are updated causing a change
in the tilt of the surface.

The {\it coarse-grained} dynamics of global tilt changes are described
by a  master equation for the probability ($P_\rho$), that the dimer model has
tilt $\rho$,
\be
{d P_{\rho} \over d t} = - \left[ W_{{\rho-1/L}, \rho} + W_{{\rho+1/L}, \rho} \right ] P_{\rho} +
W_{{\rho},{\rho-1/L}} P_{\rho -1/L} + W_{{\rho},{\rho+1/L}} P_{\rho+1/L} \ .
\label{MasterEqn}
\ee
The rates in this master equation obey the detailed balance condition: $W_{{\rho-1/L}, \rho}/
W_{{\rho},{\rho-1/L}} = \exp[-(F(\rho-1/L)-F(\rho))]$.   The usual way of achieving this balance
which leads to normal diffusive dynamics is to partition the rates symmetrically with $W_{{\rho-1/L}, \rho} \simeq \exp[-(F(\rho-1/L)-F(\rho))/2]$
and $W_{{\rho},{\rho-1/L}} = \exp[(F(\rho-1/L)-F(\rho))/2]$\cite{reichl_book}.
Equation \ref{MasterEqn}, however,  features an  unusual form for the transition
rates between different tilt states. Namely, the rates of transitions from
higher into lower tilt  states (increasing energy transitions) are determined by the energy change alone:
\be
W_{{\rho-1/L}, \rho} = \Gamma_0 e^{-(E(\rho-1/L)-E(\rho))} \ ;
\label{barr1}
\ee
here $\Gamma_0$ is a constant.
This follows from the observation that in order to lower the tilt and increase the
energy, a system spanning loop, which is always present in a state with $\rho\neq 0$,
needs to be updated.  This form of the rates for energy-increasing transitions in conjunction with the  detailed balance condition
implies that the rates of transitions to higher tilt states (energy lowering transitions) must be
determined by the entropy change: \be W_{{\rho},{\rho-1/L}} = \Gamma_0
e^{-(S(\rho-1/L)-S(\rho))} \ . \label{barr2} \ee
The form of the transition rates that we are arguing for here, was
directly observed in numerical simulations of the three coloring model
\cite{loops_epl}, which is a close relative of the dimer model. The
two are equivalent if, in the dimer models, a weight of 2 is attached
to each loop formed by bonds that are not covered by dimers.

\figone{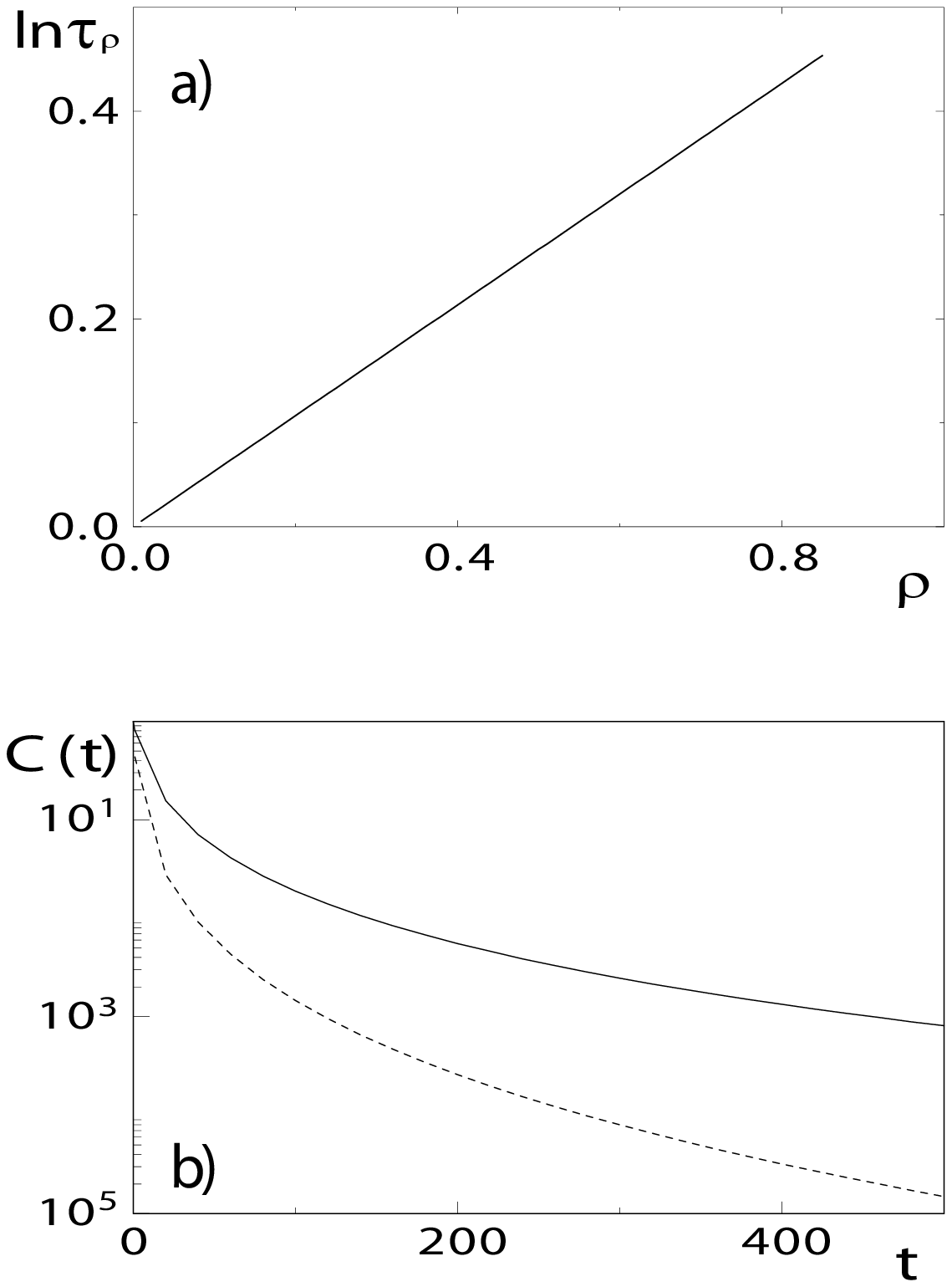}{(a) The time scales for relaxing out of
different tilt states $\rho$ in the dimer model (scaled by
$L$), for a value of $\mu$ below the transition. (b) The tilt-tilt
autocorrelation function of the dimer model. The full line is
obtained from Eq.~\ref{corr2} while the dashed line is a result of
the saddle point evaluation of Eq.~\ref{corr2}. Here $L=4096$ and
time is  measured in units of $\Gamma_{0}^{-1}$. }{Tau_fig}

\section{Relaxation time-scales}
The first consequence of the above form of the  transition rates
is that the time scale of relaxation out of a state with tilt
$\rho$, $\tau_{\rho} = 1/(W_{{\rho-1/L}, \rho} + W_{{\rho+1/L},
\rho})$, is a {\it non-decreasing} function of $\rho$. The exact
expressions for $\tau_{\rho}$ (measured in units of
$\Gamma_{0}^{-1}$),
\be
{\tau_{\rho}}^{-1} =
    { {e^{-{{16 \over 3} {\rho}\mu L}}+
e^{-L[{2 \over 3}{{\rm ln}2}+{2 \over 3}{{\rm ln}[\cos({\pi \over
3} - {{\pi {\rho}} \over 3})]}]}}} ,
\label{tauD1}
\ee
follows from Eqs.~\ref{barr1} and~\ref{barr2},
and it is plotted in  Fig.~\ref{Tau_fig}a). This
time-scale increases monotonically with $\rho$\cite{foot1}, as in
the hierarchical models of Palmer {\em et al.}~\cite{palmer}.
It is  in sharp contrast with canonical Langevin
dynamics around the equilibrium state, for which the time to relax
out of a macro-state {\em decreases} the further the  order
parameter is away from its equilibrium value. (For example, in the
Ising model with Glauber dynamics and in the disordered phase,
the  relaxation time out of a given magnetization state {\it
decreases} with increasing magnetization.)


\section{Autocorrelation function}
To quantify the tilt dynamics  we compute the tilt-tilt
autocorrelation function $C(t)$, defined as: \be C(t) = {{\langle
\rho(t) \rho(0) \rangle - {\langle \rho(0) \rangle}^{2}} \over
{\langle {\rho(0)}^2 \rangle - {\langle \rho(0) \rangle}^{2}}},
\label{corr1} \ee with the average taken over different histories
of $\rho$. An approximate form for the autocorrelation function
is: \be C(t) \approx {\sum_{\rho}{(\rho - {\langle \rho
\rangle})}^2 e^{-F(\rho)} e^{-t/{\tau_{\rho}}} \over
{\sum_{\rho}{(\rho - {\langle \rho \rangle})}^2 e^{-F(\rho)}}} ,
\label{corr2} \ee i.e., $C(t)$ is an equilibrium weighted average
of relaxations out of different $\rho$ states. This approximation
is based on the assumption that eigenfunctions of the rate matrix
are localized in $\rho$-space and eigenfunctions corresponding to
different eigenvalues do not have significant overlaps. We will
justify this assumption {\it a posteriori} by examining the
eigenfunctions obtained from numerical diagonalizations of the
rate matrix $W_{\rho,\rho'}$.
The asymptotic decay of the autocorrelation functions can be
extracted by performing a saddle point analysis of the sum in
Eq.~\ref{corr2} and using a quadratic approximation for the
entropy (Eq.~\ref{entropyD1}).  These saddle point solution is
compared in Fig.~\ref{Tau_fig}b) to the result obtained from the
sum (Eq.~\ref{corr2}).

In the limit of $\mu\to \mu_*$ and $t\to \infty$, saddle point
analysis yields:
\be
C(t) \sim
\exp\{-{3 \over 32}({{\mu_* - \mu} \over {\mu_*}^2})[{\rm ln}({{2 \mu_* t} \over {\mu_* - \mu}})]^2\} ,
\label{corrD1q} \ee
showing  that $C(t)$ in the dimer model has a log-normal form implying a  slower than exponential decay.  From Eq.~\ref{corrD1q} we also conclude that the relaxation
timescale, $\tau$, for the decay of $C(t)$ to an arbitrary
constant $C_0$, diverges exponentially as $\mu \rightarrow \mu_*$. 
This is a Vogel-Fulcher type behavior (since $\mu$ is proportional 
to $\beta = 1/{k T}$)
observed in many fragile glass formers.  First order corrections
to Eq.~\ref{corrD1q} lead to an even more rapid increase of time
scales, with ${\tau}/{\rm ln}{\tau}$ diverging as Vogel-Fulcher.

\figone{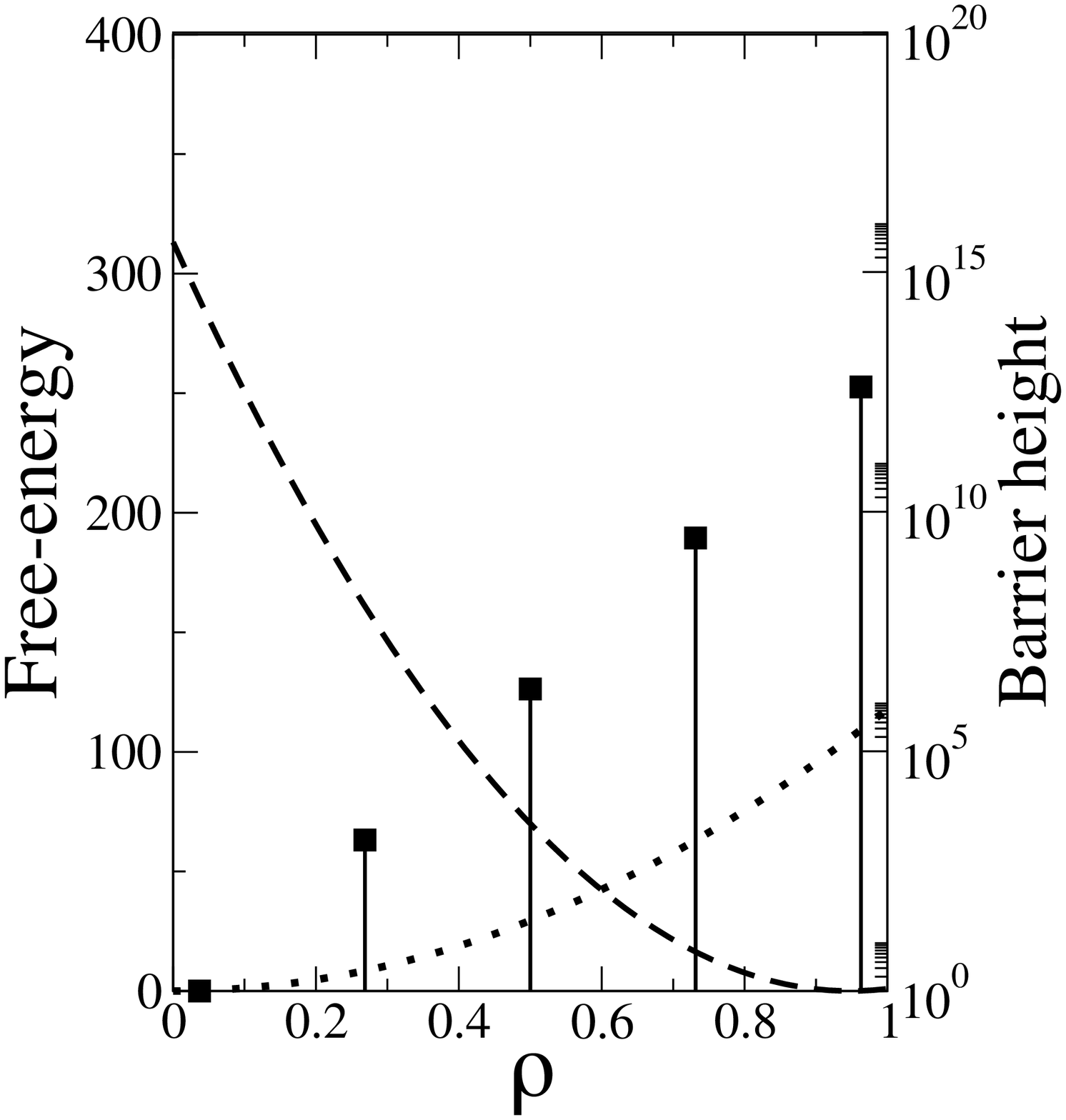}{Barrier height, $B(\rho)$ (dimensionless) shown as a set of solid
lines, and the quadratic approximation to the dimensionless free energies of the
dimer model (dashed line); $\mu$ is chosen close to ${\mu}_*$ and $L = 24$. Note the
logarithmic scale for the barrier height. }{barrier_fig}

The coarse grained dynamics defined by the transition matrix
elements, Eqs.~\ref{barr1} and~\ref{barr2}, were argued to follow
from the nonlocal loop dynamics of the dimer models. From this
form of the $W$-matrix all the conclusions about critical dynamics
of the dimer model are derived.  We have confirmed this
picture in considerable detail in simulations of the  three
coloring model \cite{messina,Kolkata}, which, as discussed
earlier, is the loop weighted dimer model. The loop weights are
not expected to affect the qualitative features of the energy and
entropy functionals. Indeed, the measured $\tau_\rho$ for the
three-coloring model compare very well
\cite{messina,Kolkata} to the analytical form plotted in
Fig.~\ref{Tau_fig}.  The numerical evidence for Vogel-Fulcher type
divergence of the relaxation time scale in this model was
reported previously \cite{messina}.

The dynamical behavior of the dimer model can be traced
back to the interplay between the free energy and  dynamical
barriers. The transition rates presented in Eqs.~\ref{barr1} and
\ref{barr2}, can be interpreted in terms of a barrier\cite{palmer}
$B({\rho}) = e^{(S(\rho-1/L)-S(\rho)+(E(\rho-1/L)-E(\rho)))/2}$
dividing the usual Metropolis rates defined in terms of the free
energy:
\be
W_{{\rho-1/L}, \rho} = \Gamma_0
e^{-(F(\rho-1/L)-F(\rho))/2}/B({\rho}) \;
\label{barrier1}
\ee
and
\be
W_{\rho, \rho-1/L} = \Gamma_0
e^{(F(\rho-1/L)-F(\rho))/2}/B({\rho}) \
\label{barrier2}
\ee
The barriers increase exponentially with $\rho$ as illustrated in
Fig.~\ref{barrier_fig}. Dynamics of the order parameter can be
viewed as relaxation in the free energy well in the presence of
these barriers.

\section{Scaling analysis of the master equation}

The emergence of the Vogel-Fulcher law in the dimer model, based on  the
master equation with transition rates defined in
Eqs.~\ref{barrier1} and \ref{barrier2}, follows from a scaling
argument.

At the critical point, $\mu = {\mu}^*$, the free energy difference
between different tilt states close to $\rho =0$ vanishes.  In
this limit, the transition rates are symmetric and given by:
\be
W_{\rho -1/L, \rho} = W_{\rho , \rho -1/L}= \Gamma_0 /B({\rho}) .
\label{symmetric}
\ee
The diffusion constant in this symmetric case can be shown to be
given by \cite{Derrida,Zwanzig}:
\be
D(L) = L \Gamma_0 /({\sum}_{i = -L,L}B({\rho}_i))
\label{Dsum}
\ee
where $L$ is the system size and ${\rho}_i  = i/L$. The longest
timescale in the problem is given by
\be
\tau (L) = L^2 /D(L)
\ee

In the limit of large $L$, the summation in Eq.~\ref{Dsum} can be replaced by an
integral. If we make use of the quadratic approximation to the entropy (Eq.~\ref{entropyD1}) then
$B(\rho ) = e^{(8/3)(\mu + {\mu}_* )L \rho }$ and the integral can be evaluated analytically, with the
result:
\be
\tau (L) = {L \over \Gamma_0} \left( e^{{16 \over 3} \mu_* L}-1 \right )
\ee
Note that the same result can be obtained by replacing the summation
by the largest barrier which occurs at $\rho =1$. The longest
time-scale in the system is, therefore, seen to diverge {\it
exponentially} with system size.  If all the barriers were equal to one,
then we would have $D(L) = \Gamma_0$ and  $\tau(L) = L^2/ \Gamma_0$, which
corresponds to simple diffusion.  In the presence of the barriers, $D(L)$ goes to
zero exponentially and this leads to an exponential divergence of the relaxation time-scale.

We can now use a scaling argument to deduce the behavior of the relaxation time-scale for $\mu
< {\mu}_*$.  The effective scale (in $\rho$-space) over which the free-energy
well is flat, and therefore the transition rates are symmetric,
diverges as the phase transition at  $\mu = {\mu}_*$ is approached. We argue that this
length scale, given by $l(\mu) = {{\sqrt{\mu}_*} \over
{\sqrt{{\mu}_* -\mu}}}$, provides a cutoff to the
summation (or integral)  involved in calculating the diffusion constant
\be
D(l) = \Gamma_{0}l/({\sum}_{i = -l(\mu ),l (\mu )}B({\rho}_i))
\ee
The longest time scale therefore scales as
\be
\tau(\mu)  = \tau (l(\mu)); \  \  l(\mu)<L
\ee
and
\be
\tau(\mu)  = \tau (L);   \  \ l(\mu) \geq L \ .
\ee
In the thermodynamic limit, $\tau (\mu )$ diverges as $\tau (\mu )
\simeq e^{(16/3){{\sqrt{\mu}_*} \over {\sqrt{{\mu}_* -\mu}}}}$.   Since  $\mu \simeq 1/T$,  $\tau$ has
a Vogel-Fulcher type divergence $\tau (T) \simeq  e^{(16/3){{\sqrt{T}_*} \over {\sqrt{T -T_{*}}}}}$.

We have recently shown that the Vogel-Fulcher divergence of the
dimer model can be obtained from an exact solution of the continuum
version of the master equation if we assume that $S(\rho)$ has a
quadratic form \cite{satya_rapid}.

\section{Numerical analysis of the master equation}

We have carried out numerical diagonalizations of the rate matrix
for $L{\rm x}L$ systems in order to verify  some of the
assumptions that have been made in the scaling analysis and  in the
calculation of the correlation function. These computations also provide us with information
about the finite-size effects on the critical dynamics of the dimer model.

The probability distribution, $P(\rho, t)$,  can be written in
terms of the eigenvalues, $\lambda_i$, and eigenfunctions, $\psi_i
(\rho)$ of the rate matrix\cite{reichl_book}:
\begin{equation}
P(\rho, t) = {\sum}_i \psi_i (\rho) e^{-{\lambda}_i t}
\label{eigen}
\end{equation}

The eigenvalues of the rate matrix are non-negative and the
equilibrium distribution is given by the zero-eigenvalue function,
${\psi_1}$ ($\lambda_1 =0$).  The smallest, non-zero eigenvalue characterizes the
state with the longest relaxation time.  All correlation functions
can be expressed in terms of the eigenvalue spectrum and, in
particular, the equilibrium, tilt-tilt autocorrelation function
can be written as:
\begin{equation}
C(t) = {\sum}_i e^{-\lambda_i t} {\sum}_{\rho \rho^\prime} \rho
\rho^\prime e^{-(F_{\rho} + F_{\rho^\prime})/2}
\psi_i(\rho) \psi_i^*(\rho^\prime) \label{autocorr_eigen}
\end{equation}
Comparing to Eq.~\ref{corr2}  it follows that the approximate
form is obtained in the limit of delta-function localized
eigenfunctions.  We will show below that the eigenfunctions
corresponding to non-zero eigenvalues of the dimer model are indeed
well localized.

Results of the  numerical diagonalization, using the exact
entropy function, show that the longest time scale, $\tau$,  of the dimer model
increases in a non-arrhenius, Vogel-Fulcher fashion, as shown in
Fig.~\ref{scaled_fig}.  The scaling in this figure is what is
expected from the scaling solution of the model with the quadratic
entropy \cite{satya_rapid} and similar to the results obtained from
the scaling arguments presented in this paper, however the length
scale emerging is $l(\mu)=[{\mu^* \over {\mu^* - \mu}}]^2$ not
$l(\mu) = {{\sqrt{\mu}_*} \over {\sqrt{{\mu}_* -\mu}}}$, as would
be expected from the quadratic entropy results \cite{satya_rapid}.

\figone{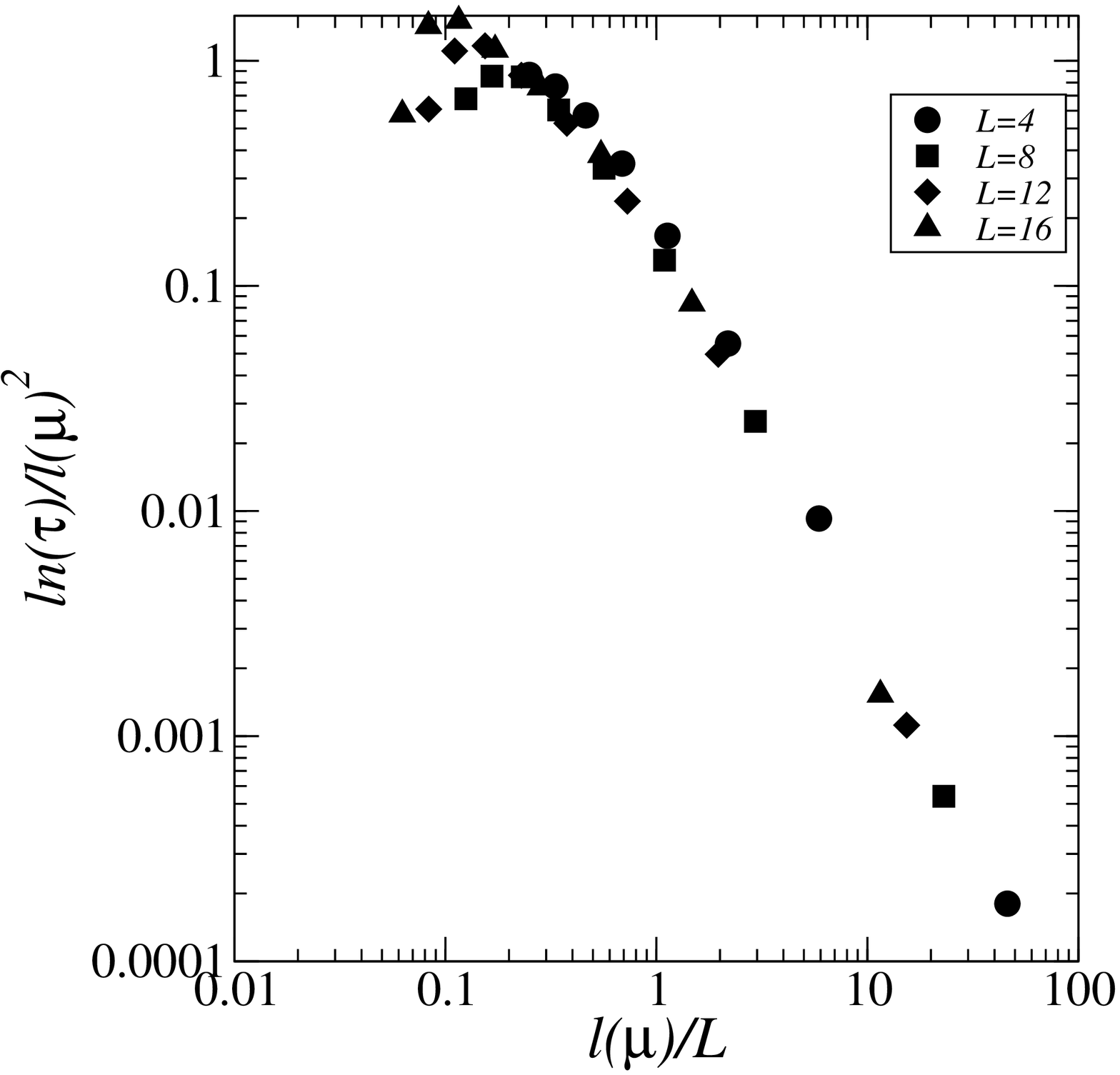}{Scaling of $\tau(\mu,L)$ in the dimer
model. The figure shows that a scaling form can be constructed in
terms of the lengthscale $l(\mu)=(\mu^* /(\mu^* - \mu))^2$ and the
particular form demonstrates the Vogel-Fulcher scaling ${\tau}
\simeq e^{(l(\mu))^2}$.}{scaled_fig}

The exact
diagonalization results show, unambiguously, that  the Vogel-Fulcher law
characterizes the time scale divergence at the entropy vanishing
transition in the dimer model.

\subsection{Eigenfunctions}
In the dimer model, the eigenfunction corresponding to the
smallest non-zero eigenvalue is localized at the largest barrier,
 i.e, the largest value of $\rho$. Higher eigenfunction move
to smaller barriers but are still localized.  The expression we
used for $C(t)$ is exact for delta-function localization of
eigenfunctions and the numerical results justify this assumption,
{\it a posteriori}.
\figone{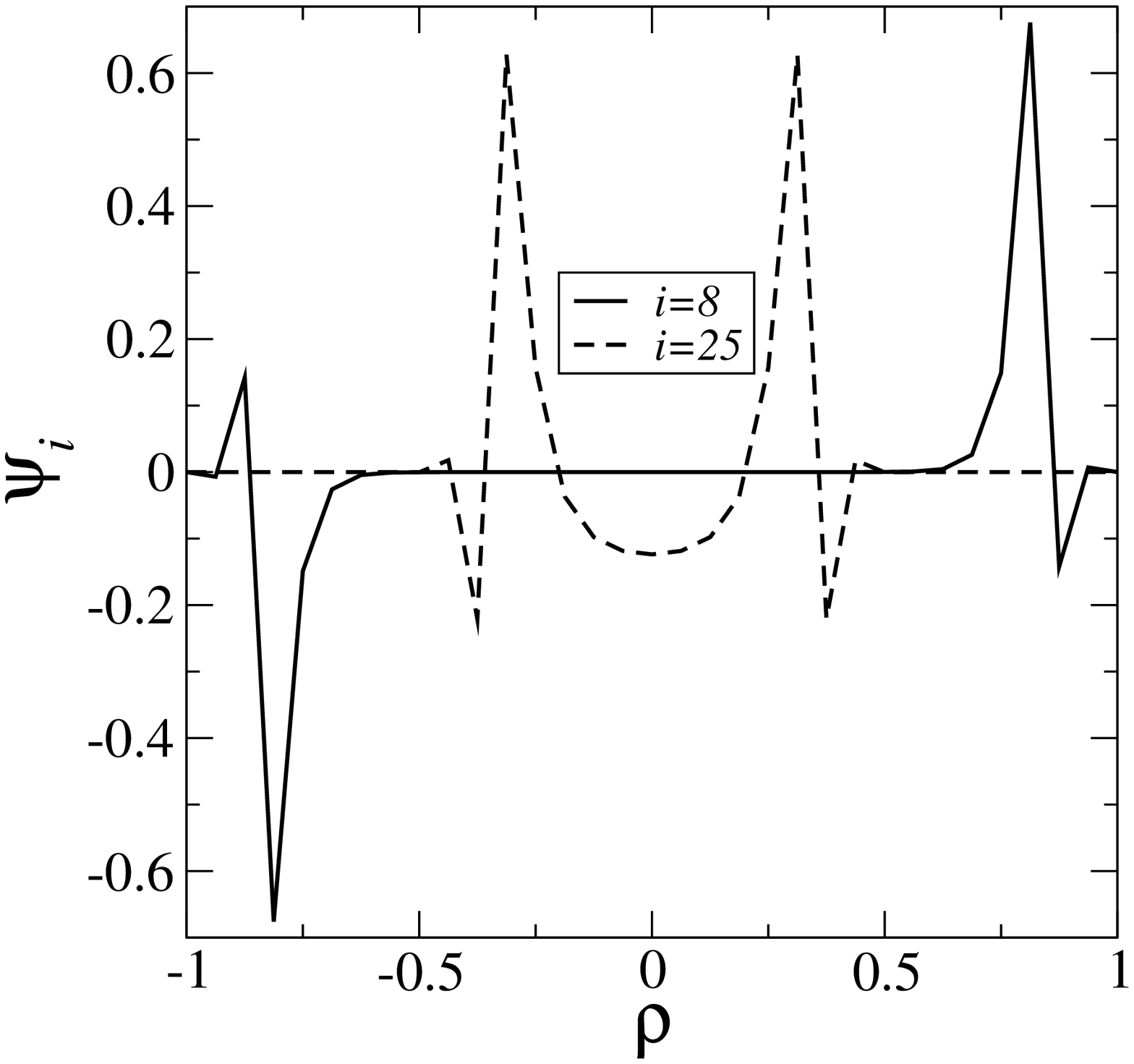}{Plots of the eighth and twenty-fifth
eigenfunctions of the dimer model for L=16 and $\mu \simeq \mu*$. These two eigenfunctions
were chosen to illustrate the localization of the eigenfunctions
and the shift towards $\rho =0$ with increasing spectral
index}{eigenfn_fig}

\subsection{Sensitivity of dynamics to barrier size}
The sensitivity of the relaxation times to the barriers heights has
been  investigated by using the quadratic entropy model and
writing $B({\rho}) = e^[{c \over 2}
{(S(\rho-1/L)-S(\rho)+(E(\rho-1/L)-E(\rho)))/2}]$ and varying $c$
between $0$ and $1$.  For $c=0$, we recover the usual Langevin
dynamics and the time scales should increase as a power law and
for $c=1$ we have barriers corresponding to the loop dynamics. The
results plotted in Fig \ref{barrier_strength} demonstrate that,
for $c=0$, $\tau \simeq (\mu^* - \mu)^{-1}$ which is consistent
with a dynamical exponent $z=2$ and a correlation length exponent
of $\nu = 1/2$; the exponents expected from a Langevin description
of a mean-field model.  It is also clearly seen from this figure
that even for $c=0.25$, the timescale increases more rapidly than
a power law.   An analysis of the continuum limit of the dimer
model dynamics  shows that, for any non-zero
value of $c$,  $\tau \simeq e^{A(c^{2}l(\mu))}$ where $A$ is a constant \cite{satya_rapid,dibunpub}.
These results taken all together imply that there is a
whole class of systems, where dynamical constraints may lead to non-zero values
of $c$, which belong to different universality classes of
dynamical critical phenomena. These are characterized by a Vogel-Fulcher rather than a
power-law divergence of relaxation time-scales, with $c$ being an
indicator of fragility \cite{angel_review}.  In real systems, such as
supercooled liquids, one expects that there is a large but finite
energy scale at which the hard constraints are violated. This energy scale then leads to a
long-time cutoff of the Vogel-Fulcher behavior.  This time
scale, may however, be well beyond any experimentally measurable
time scales.

\figone{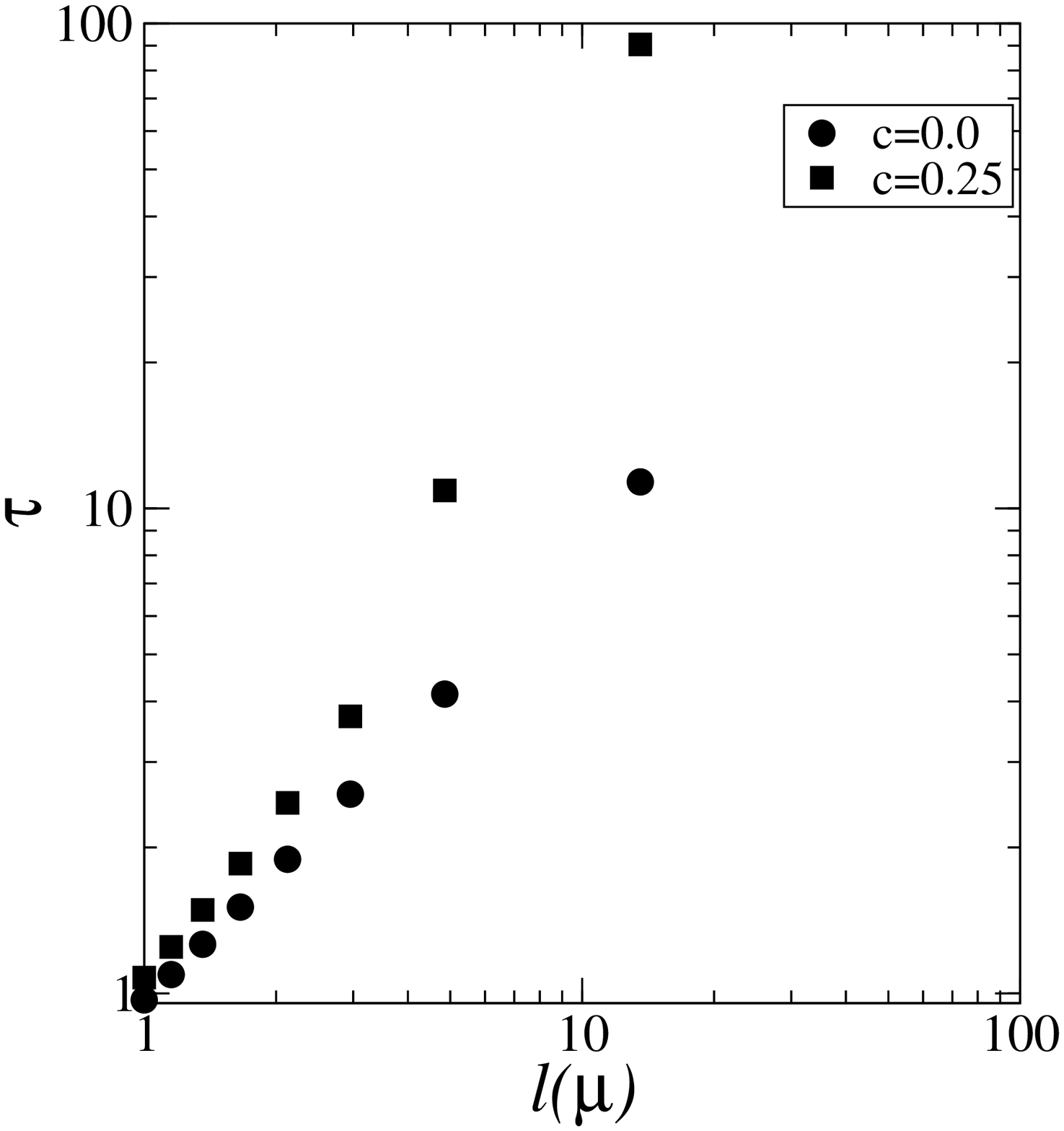}{Timescale $\tau$ in the dimer model for
different values of the barrier strength, $c$, plotted as a
function of $l(\mu) = {\mu^* \over {\mu^* - \mu}}$}{barrier_strength}


\section{Adam-Gibbs scenario}


The entropy of the dimer model, $S_{conf}$,
which corresponds to $S(\rho)$ evaluated at the equilibrium value of the order parameter $\rho$,
goes to zero at the transition. Furthermore, our results clearly show that the longest time
scale diverges in a Vogel-Fulcher manner.   The Adam-Gibbs
relation, $\tau = {\tau}_0 e^{A/{S_{conf}(T)}}$, however, does not
capture the physics since in the dimer model,  $S_{conf}$ jumps from a finite value at $\rho =0$ to
zero, as $\rho$ changes discontinuously to $1$.
Thus, in this model, the exponential divergence of the relaxation time-scale at the transition   is not
accompanied by a continuous vanishing of $S_{conf}$.
The analysis
presented in this paper clearly demonstrates that the
Vogel-Fulcher divergence is rooted in the constrains which lead to loop dynamics.   This type of nonlocal dynamics leads to
 the unusual transition rates with energy-lowering transitions being determined by
changes in entropy and, therefore, to an exponential decrease of
the number of energy-lowering trajectories as one approaches the
zero-entropy state.

The configurational entropy of supercooled liquids has been
interpreted as the inherent structure entropy, i.e., the number of
valleys at the temperature of interest.  The observation that the
Adam-Gibbs scenario describes much of the phenomenology of
supercooled liquids could imply that there is a phase transition in
the inherent-structure space similar to the one discussed in this paper
for dimers.   Experiments and simulations have shown that a hallmark of supercooled liquids approaching the glass
transition is the appearance of dynamical heterogeneities\cite{weitz,ediger,glotzer}.  The loops in the dimer
dynamics are  analogs of these dynamical heterogeneities since they define the correlated moves allowed by the constraints.
These heterogeneities  are present as long as the constraints are not violated and are characterized by a size distribution
which changes as the critical point is approached\cite{Kolkata,messina}.   The analogy between loops and dynamical heterogeneities
suggest that the dynamics in the inherent structure space of supercooled liquids could be similar to the loop dynamics of dimer models.
If this is the case then transition rates between inherent structures should exhibit features similar to the ones discussed in this paper.
We
are currently in the process of  analyzing
transition rates between  inherent structures of
Lennard-Jones glass formers in order to get a better understanding of the connection between dynamical heterogeneities and the effective
dynamics  in the inherent-structure space.

The authors would like to acknowledge numerous useful discussions
with Satya Majumdar.  This work was supported by grants from the
NSF: DMR-0207106 (DD and BC),
DMR-9984471 (JK) and
DMR-0403997 (BC and JK). JK is a Cottrell Scholar of Research Corporation.


\begin{thebibliography}{0}
\bibitem{foot2}The components of the tilt along the three lattice
directions of the triangular lattice are subject to two constraints.
One arising from the conservation of the total dimer number ($= L^2)$
and the other from periodic boundary conditions which force two of the
components to be equal.
\bibitem{foot1}Since the phase transition occurs along the metastable
branch in the dimer model, the range of ${\rho}$'s participating in the
dynamics is restricted to ${\rho} \le {\rho}_{max}$. The free energy has
a maximum at ${\rho}_{max}$ which vanishes as $\mu \rightarrow {\mu}_*$
and scales with the system size.
\end{thebibliography}
\end{document}